\documentclass[a4paper,11pt]{article}

\usepackage{amssymb}
\usepackage{amsmath}
\usepackage{graphicx}
\usepackage{comment}

\begin{document}

\newcommand{\bin}[2]{\left(\begin{array}{c} \!\!#1\!\! \\  \!\!#2\!\!
\end{array}\right)}
\newcommand{\troisj}[3]{\left(\begin{array}{ccc}#1 & #2 & #3 \\ 0 & 0 & 0
\end{array}\right)}
\newcommand{\sixj}[6]{\left\{\begin{array}{ccc}#1 & #2 & #3 \\ #4 & #5 & #6
\end{array}\right\}}
\newcommand{\neufj}[9]{\left\{\begin{array}{ccc}#1 & #2 & #3 \\ #4 & #5 & #6 \\
#7 & #8 & #9 \end{array}\right\}}
\newcommand{\der}[2]{\frac{\partial #1}{\partial #2}}
\newcommand{\fpt}{\mathcal{T}}
\newcommand{\fpf}{\mathcal{F}}
\newcommand{\fpg}{\mathcal{G}}
\newcommand{\fpn}{\mathcal{N}}
\newcommand{\fpr}{\mathcal{R}}
\newcommand{\fpu}{\mathcal{U}}
\newcommand{\fpv}{\tilde{\mathcal{V}}}
\newcommand{\fpvv}{\mathcal{V}}
\newcommand{\fpw}{\mathcal{W}}
\newcommand{\fpz}{\mathcal{Z}}
\newcommand{\fph}{\mathcal{H}}
\newcommand{\pinE}{\vec p\;\in\;\mathcal{E}}
\newcommand{\calE}{\mathcal{E}}

\title{Jensen-Feynman approach to the statistics of interacting electrons}

\author{Jean-.Christophe Pain$^{\dag}$\footnote{phone: 00 33 1 69 26 41 85, email: jean-christophe.pain@cea.fr}, Franck Gilleron$^{\dag}$ and G\'erald Faussurier\footnote{CEA/DIF, B.P. 12, 91680 Bruy\`eres-Le-Ch\^atel Cedex, France}}

\maketitle


\begin{abstract}
Faussurier \emph{et al.} [Phys. Rev. E \textbf{65}, 016403 (2001)] proposed to use a variational principle relying on Jensen-Feynman (or Gibbs-Bogoliubov) inequality in order to optimize the accounting for two-particle interactions in the calculation of canonical partition functions. It consists in a decomposition into a reference electron system and a first-order correction. The procedure appears to be very efficient in order to evaluate the free energy and the orbital populations. In this work, we present numerical applications of the method and propose to extend it using a reference energy which includes the interaction between two electrons inside a given orbital. This is possible thanks to our efficient recursion relation for the calculation of partition functions. We also show that a linear reference energy, however, is usually sufficient to achieve a good precision and that the most promising way to improve the approach of Faussurier \emph{et al.} is to apply Jensen's inequality to a more convenient convex function.
\end{abstract}

\section{Introduction}

The superconfiguration method \cite{bar} is a powerful technique for the study of opacity and equation of state of hot plasmas. The formalism requires the calculation of
the partition functions of the superconfigurations using recursive methods, and relies on independent-particle statistics. For
that purpose, in the original STA (Super Transition Arrays) approach, two-particle interactions are averaged over the
configurations of a given superconfiguration. The formalism proposed by Faussurier \emph{et al.} \cite{fau} consists in an optimization of the orbital energies taking into account the interactions using a variational principle.  This method reduces drastically the number of superconfigurations required for the convergence of STA calculations, due to a better treatment of the electron correlations. In the present work, we investigate two possible improvements of the latter approach. 

The first one consists in using a reference energy which is quadratic with respect to the populations. This enables one to take into account the interaction between two electrons belonging to the same orbital. Moreover, it can be easily understood that a quadratic reference energy is more appropriate for the description of a system including quadratic two-body interactions. We show, however, using a more general reference energy, that the linear reference energy gives results that are already very close to the exact values. 

The second improvement consists in applying Jensen's inequality to another convex function \cite{dec}, which is the difference between the exponential function and its $(2n-1)^{th}$-order Taylor development. Such a procedure requires the computation of high-order moments, but appears to be very powerful.

The paper is organized as follows. In Sec. \ref{sec2}, we express the partition function of interacting electrons with an arbitrary reference system that can be optimized through a variational principle. In Sec. \ref{sec3}, we study the impact of choosing a particular reference energy on the accuracy of the method. We discuss the possibility, in our new recursion relation \cite{gil,wil}, to use a non-linear reference energy, in order to better account for two-electron interactions. In Sec. \ref{sec4}, we propose an improved Jensen-Feynman inequality. Sec. \ref{sec5} is the conclusion.

\section{\label{sec2}Statistics of interacting electrons}

\subsection{Generalities}

The central-field model is the simplest approach for calculating the atomic structure of an interacting many-electron system. Each electron is assumed to move independently in a central potential that represents the electrostatic field of the nucleus and the spherically averaged mutual repulsions of the other electrons. This central-field Hamiltonian allows one to characterize the quantum states of the system (energies and wavefunctions) by means of degenerate relativistic or non-relativistic electron configurations. The latter are defined as groups of degenerate subshells of the type $(n_i\ell_i)^{p_i}$ (non-relativitic case) or $(n_i\ell_i j_i)^{p_i}$ (relativistic case), where  $n_i$ is the principal quantum number, $\ell_i$ is the orbital quantum number, $j_i=\ell_i\pm 1/2$ the relativistic angular momentum, and $p_i$ is the population of the $i^{th}$ orbital such that $0\le p_i\le g_i$ (the degeneracy $g_i$ is equal to $4\ell_i+2$ in the non-relativistic case, and to $2j_i+1$ in the relativistic case). Since configurations differ only by the electron populations of orbitals, it is useful to represent them by the vector $\vec p=\{p_1,p_2,\ldots p_N\}$ of $N$ given orbitals.
The accuracy of the central-field model can then be improved by evaluating the non-central parts of the Hamiltonian using the first-order perturbation theory. Accounting for one- and two-body operators in the Hamiltonian of the system, the energy $E(\vec p)$ of a configuration reads

\begin{eqnarray}\label{energy}
E(\vec p)&=&\sum_{i=1}^N p_i\epsilon_i+\frac{1}{2}\sum_{i,j=1}^N p_i(p_j-\delta_{ij})V_{ij},
\end{eqnarray}

where $\epsilon_i$ (energy of orbital $i$) and $V_{ij}$ (interaction energy between orbitals $i$ and $j$) are evaluated from the central-field Hamiltonian, and $\delta_{ij}$ is Kronecker's symbol.

The main purpose of this paper is to study the statistics of physical quantities over an arbitrary canonical ensemble (denoted $\mathcal E$) of configurations, with $Q$ electrons populating $N$ orbitals. The set of configurations in $\calE$ is obtained by fixing or varying the population of each orbital, with the constraint $\sum_{i=1}^Np_i=Q$. The average value of any quantity $A(\vec p)$ over all the configurations of $\calE$ reads

\begin{eqnarray}\label{def2:moyA}
\left\langle A\right\rangle=\frac{1}{U(\calE)}\sum_{\pinE}A(\vec p)~G(\vec p)~e^{-\beta E(\vec p)}.
\end{eqnarray}

In this expression, $U(\calE)$ is the partition function of the ensemble $\mathcal E$ and is defined as

\begin{eqnarray}\label{def:fp}
U(\calE)=e^{-\beta F}=\sum_{\pinE}G(\vec p)~e^{-\beta E(\vec p)},
\end{eqnarray}

where $F$ is the total free energy and $G(\vec p)$ the degeneracy of the configuration $\vec p$:

\begin{eqnarray}
G(\vec p)=\prod_{i=1}^N\bin{g_i}{p_i}.
\end{eqnarray}

Such expression of the degeneracy involving binomial factors is the signature of the Fermi statistics, \emph{i.e.} of the proper accounting for the Pauli exclusion principle. It was shown in \cite{fau2}, using an integral representation of the partition function evaluated with the saddle-point method, that the average populations of orbitals obey Fermi-Dirac distribution in the limit of a large number of electron states.

The sums contained in formulas (\ref{def2:moyA}) and (\ref{def:fp}) run usually over a very large number of electron configurations. A direct evaluation of these expressions is therefore a hard and sometimes an almost impossible task.
Unfortunately, the quadratic dependence with respect to the populations of orbitals (due to two-body interactions) is known to prevent any factorization of the partition function. This drawback makes impossible the use of recursion techniques to speed up the calculation of averages.

\subsection{Use of a reference system}

A way to avoid this factorization issue is to introduce an arbitrary reference energy $E_R(\vec p)$, writing:

\begin{eqnarray}\label{eqint}
U(\calE)&=&\sum_{\pinE}G(\vec p)~e^{-\beta [E_R(\vec p)+\Delta E(\vec p)]}\\
&=&\left\langle e^{-\beta \Delta E}\right\rangle_R ~U_R(\calE)\label{eqint2},
\end{eqnarray}

with $\Delta E=E(\vec p)-E_R(\vec p)$. The letter \textit{R} means that the assigned quantity is evaluated in the new reference system by replacing $E(\vec p)$ by $E_R(\vec p)$. The notation $\langle A\rangle_R$ is thus defined as

\begin{eqnarray}\label{moyA}
\left\langle A\right\rangle_R=\frac{1}{U_R(\calE)}\sum_{\pinE}A(\vec p)~G(\vec p)~e^{-\beta E_R(\vec p)},
\end{eqnarray}

where

\begin{eqnarray}\label{def:fpr}
U_R(\calE)=e^{-\beta F_R}=\sum_{\pinE}G(\vec p)~e^{-\beta E_R(\vec p)}.
\end{eqnarray}

The mean value of any quantity in the genuine interacting system can now be put in the form:

\begin{equation}\label{moy}
\left\langle A\right\rangle=\frac{\left\langle A e^{-\beta\Delta E}\right\rangle_R}{\left\langle e^{-\beta\Delta E}\right\rangle_R}
\end{equation}

which depends only on averages in the new reference system.

The choice of the reference energy $E_R(\vec{p})$, though arbitrary, must enable one to evaluate $U_R(\calE)$ using recursion relations (factorizability). This is discussed in Sec. III. Moreover, only averages of quantities which are multinomial functions of the populations (\emph{e.g.} $p_i$, $p_i^3p_j$, $p_i^4p_j^2p_k^3$, \emph{etc.}) can be expressed in terms of partition functions (see Appendix A). Therefore, an approximation has to be made for the exponential term in Eq. (\ref{moy}). This is discussed in the following subsection. 

\subsection{Variational approach based on the Jensen-Feynman inequality}

Jensen's inequality \cite{jen} states that for any continuous and convex function $u\rightarrow f(u)$, one has

\begin{equation}\label{eqj}
\left\langle f(u)\right\rangle\geq f(\left\langle u\right\rangle),
\end{equation}

which leads, for $f(u)=e^u$, to

\begin{equation}
\left\langle e^{-\beta\Delta E}\right\rangle_R\geq e^{-\beta\left\langle\Delta E \right\rangle_R}
\end{equation}

and therefore from Eq. (\ref{eqint2})

\begin{eqnarray}
U(\calE)\ge e^{-\beta\left\langle\Delta E\right\rangle_R}~U_R(\calE).
\end{eqnarray}

Taking the natural logarithm of this expression, we obtain the first-order Jensen-Feynman (or Gibbs-Bogoliubov) \cite{fey,bog,isi} inequality:

\begin{eqnarray}\label{bogo}
F \le F^{(1)},
\end{eqnarray}

where 

\begin{equation}\label{fr1}
F^{(1)}=F_R +\left\langle\Delta E\right\rangle_R
\end{equation}

is an upper-bound approximation of the thermodynamic potential $F$. If the reference energy contains free parameters, they can be optimized through a minimization procedure, providing in that way a better approximation for $F$. 

It can be noticed that the Jensen-Feynman approach for deriving $F^{(1)}$ is also consistent with the approximation $e^{-\beta[\Delta E-\left\langle\Delta E \right\rangle_R]}\simeq 1$ in the expression of $F$. Performing the same approximation in Eq. (\ref{moy}) leads to

\begin{eqnarray}\label{moyR}
\left\langle A\right\rangle\simeq\left\langle A\right\rangle_R.
\end{eqnarray}

\section{\label{sec3}First prescription: Choice of the reference system}

The choice of the reference system must allow one to factorize the partition functions in order to derive recursion relations. Recently, we have proposed an efficient technique \cite{gil,wil} to compute the partition functions. It is based on a doubly nested recursion (on the number of electrons and orbitals), each orbital being added one after the other. Because of the orbital separability, this approach can be applied to a reference system with an energy of the form $E_R(\vec p)=\sum_{i=1}^N \zeta_i(p_i)$, where $\zeta_i$ is an arbitrary function of the population of orbital $i$. Partition functions are then derived from the efficient recursion relation

\begin{equation}
U_{Q;N}^R=\sum_{p_N=0}^{\min(Q,g_N)}~e^{-\beta\zeta_N}U_{Q-p_N;N-1}^R,
\end{equation}

initialized with $U_{Q;0}^R=\delta_{Q,0}$. An important constraint of the recursive techniques is that only the average values of quantities containing integer powers of the populations can be deduced from the knowledge of partition functions (see Appendix A).

In this section, we study the impact of choosing a particular reference energy on the accuracy of the method. A non-exhaustive list of reference energies is presented in Table \ref{tab2bis}.

In the reference energy A, the quadratic terms with respect to the populations are averaged over all the configurations. This is the method used in the original superconfiguration theory \cite{bar}. The case B corresponds to the work of Faussurier \emph{et al.}: it consists of a linear form of the populations and has $N$ degrees of freedom (namely $\{\theta_i, i=1,N\}$) obtained by minimizing the right-hand side of Eqs. (\ref{bogo}). The reference energy C is an extended version of case B where a quadratic dependence of the populations in a given orbital is added.

The case D is a power-law form which is introduced to infer some information about the optimal exponents in the reference energy. It requires $2N$ parameters $\{\theta_i,\gamma_i,\; i=1,N\}$. It can be noticed that this choice prevents any closed-form evaluation of $<\Delta E>_R$ in terms of recursive partition functions, because $E_R(\vec{p})$ is not a multinomial function. In this case, the calculations are performed "brute force" by summing over all the configurations.

In case A (no adjustable parameters), the free energy is directly obtained by evaluating Eq. (\ref{fr1}). In cases B, C and D, the free parameters are determined by minimizing the same equation.
In practice, the minimization of the free energy $F^{(1)}$ is performed with a conjugate-gradient method by using
the vector of derivatives with respect to the $K$ free parameters $\{a_i, i=1,K\}$:

\begin{equation}
\vec{\nabla}F^{(1)}=\left\{\begin{array}{cccccccc} 
\der{F^{(1)}}{a_1}, & \ldots & ,\der{F^{(1)}}{a_K}\end{array}\right\},
\end{equation}

which can be expressed analytically. For instance, for reference energy C, one has 

\begin{eqnarray}
\der{F^{(1)}}{\theta_k}&=&\beta\sum_{i=1}^N[\left\langle p_i\right\rangle_R\left\langle p_k\right\rangle_R-\left\langle p_ip_k\right\rangle_R](\epsilon_i-\theta_i)\nonumber\\
&&+\frac{\beta}{2}\sum_{i,j=1}^N[\left\langle p_i(p_j-\delta_{ij})\right\rangle_R\left\langle p_k\right\rangle_R\nonumber\\
&&-\left\langle p_i(p_j-\delta_{ij})p_k\right\rangle_R](V_{ij}-\phi_i\delta_{ij})
\end{eqnarray}

and 

\begin{eqnarray}
\der{F^{(1)}}{\phi_k}&=&\frac{\beta}{2}\sum_{i=1}^N[\left\langle p_i\right\rangle_R\left\langle p_k(p_k-1)\right\rangle_R\nonumber\\
&&-\left\langle p_ip_k(p_k-1)\right\rangle_R](\epsilon_i-\theta_i)\nonumber\\
&&+\frac{\beta}{4}\sum_{i,j=1}^N[\left\langle p_i(p_j-\delta_{ij})\right\rangle_R\left\langle p_k(p_k-1)\right\rangle_R\nonumber\\
&&-\left\langle p_i(p_j-\delta_{ij})p_k(p_k-1)\right\rangle_R](V_{ij}-\phi_i\delta_{ij}).\nonumber\\
&&
\end{eqnarray}

where the mean quantities are calculated from the partition functions of the reference system. 

For numerical illustration, we consider the case discussed in Ref. \cite{wil93}, \emph{i.e.}, a copper plasma at  $T$=100 eV and $\rho$=8.96 g/cm$^3$. The statistics is performed over all relativistic configurations of the type

$$
K^2~L^8~(3s_{1/2}~3p_{1/2}~3p_{3/2}~3d_{3/2}~3d_{5/2}~4s_{1/2}~4p_{1/2}~4p_{3/2})^Q,
$$

where $Q$ electrons (which may vary from $0$ to $26$) are distributed over the orbitals in parenthesis. The K ($n=1$) and L ($n=2$) shells are assumed to be full. The free energies presented and discussed below do not include the contribution of the frozen core. We can see in Table \ref{tab1}, for different values of the number of electrons $Q$ (first column), that the free energies when the interactions are artificially cancelled (third column) are very different from the exact values (second column). It is obvious that the inclusion of the interactions is crucial: the maximum error reaches about 65 $\%$ for $Q$=13 (half the total degeneracy). When the interactions are averaged (fourth column, reference energy A), the maximum error (still reached for $Q$=13) is reduced to 8 $\%$. It is interesting to mention that the results do not change if the reference energy is shifted by a quantity which is independent of the populations (see Appendix B). The results can be improved by minimizing the free energy. For instance, with a linear reference energy and first-order Jensen-Feynman inequality (fifth column, $F^{(1)}$ with reference energy B), the maximum discrepancy between the obtained free energies and the exact values drops to 0.03 $\%$. One finds that the use of a quadratic reference energy, which introduces twice the number of free parameters (sixth column, reference energy C) brings a slight improvement of the results which were already excellent using the reference energy B. This can be understood considering the "power-law" reference energy (seventh column, reference energy D). One finds that the exponents obtained after minimization are very close to one (see Table \ref{tab2}) for all the eight orbitals, which justifies the choice of the linear reference energy. One can also notice that the free energies obtained with the power-law reference energy D are very close to the ones obtained with the quadratic reference energy C. This is another evidence that the linear part of the energy prevails over the quadratic part.

\begin{figure}
\begin{center}
\includegraphics[width=8cm]{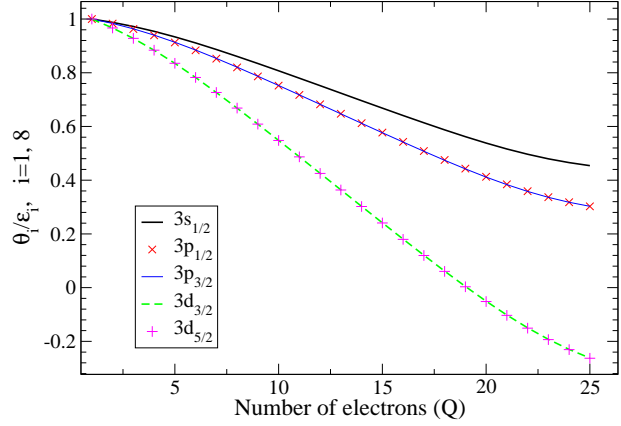}
\caption{(Color online) Ratio $\theta_i/\epsilon_i$ for orbital $i$ ($3s_{1/2}$ to $3d_{5/2}$) obtained after minimization of the free energy within first-order Jensen-Feynman inequality (\ref{bogo}), using reference energy B. The ratios for $3p_{1/2}$ (thin line) and $3p_{3/2}$ (crosses) are almost indistinguishable, as well as the ratios for $3d_{3/2}$ (dashed line) and $3p_{5/2}$ (plus symbols).}
\label{fig1}
\end{center}
\end{figure}

\begin{figure}
\begin{center}
\includegraphics[width=8cm]{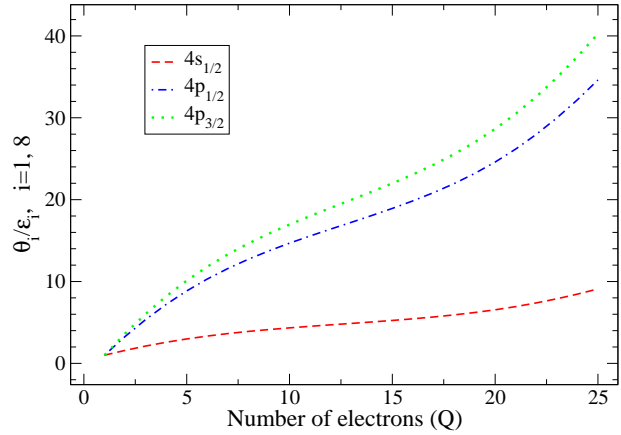}
\caption{(Color online) Ratio $\theta_i/\epsilon_i$ for each orbital $i$ ($4s_{5/2}$ to $4p_{3/2}$) obtained after minimization of the free energy within first-order Jensen-Feynman inequality (\ref{bogo}), using reference energy B.}
\label{fig2}
\end{center}
\end{figure}

Figures \ref{fig1} and \ref{fig2} display, for each orbital $i$, the ratio $\theta_i/\epsilon_i$ obtained after minimization of the free energy with reference energy B and first-order Jensen-Feynman inequality ($F^{(1)}$). One can see that for the highest-energy orbitals, $\theta_i$ can be very different from $\epsilon_i$, which can be explained by the fact that the electrons in such orbitals are very sensitive to electron-electron interactions, on the contrary to electrons in the lower orbitals which are more subject to the attraction of the nucleus.

One can be surprised that some values of $\theta_i$ become positive (orbitals $3d_{3/2}$ and $3d_{5/2}$, see Fig. \ref{fig1}) for some values of $Q$. In fact, the reference energy B is the energy of a fictious non-interacting system. It represents both the linear part of the energy of the real system and the quadratic interactions, which are positive. Therefore, when the contribution of the quadratic interactions dominates, $\theta_i$ can become positive.

The strength of the method resides in the fact that, even when $\theta_i$ differs notably from $\epsilon_i$, the populations of the corresponding orbitals are very close to the exact results, which was rather unexpected.

\section{\label{sec4}Second prescription: higher-order Jensen-Feynman inequality}

A better approximation may be found by applying Jensen's inequality (\ref{eqj}) to the convex function

\begin{equation}
f_n(u)=e^u-\sum_{k=0}^{2n-1}\frac{u^{k}}{k!}
\end{equation}

with $u=-\beta[\Delta E(\vec{p})-\left\langle\Delta E(\vec{p})\right\rangle_R]$. We obtain a new Jensen-Feynman inequality \cite{dec} with an adjustable precision driven by $n$, which reads

\begin{equation}\label{newgb}
F\leq F^{(2n-1)},
\end{equation}

where

\begin{equation}\label{fr3}
F^{(2n-1)}=F_R+\left\langle\Delta E\right\rangle_R-\frac{1}{\beta}\ln\left[1+\sum_{k=2}^{2n-1}(-1)^k\frac{\beta^k}{k!}M_k\right],
\end{equation}

and $M_k=\left\langle[\Delta E-\left\langle\Delta E\right\rangle_R]^k\right\rangle_R$. The expression of $F^{(2n-1)}$ is more and more complicated as the order $n$ increases, but all the quantities $M_k$, $k$=1,$\cdots, 2n-1$  can still be obtained using our efficient recursion relations.
The last column of Table \ref{tab1} shows the values of the free energy developed at the third order (\emph{i.e.} $n=2$) and minimized with the linear reference energy C. We can see that, even with this simple reference energy, the obtained free energy $F^{(3)}$ is always closer to the exact value than $F^{(1)}$. The precision reached by increasing the order of Jensen-Feynman inequality is better than the one obtained within changing the reference energy and/or increasing the number of free parameters. We have checked that an arbitrary precision can be obtained by going to higher order ($n$=3 or more). However, it must be clear that the computational time increases rapidly with $n$ due to multiple evaluations of $F^{(2n-1)}$ by the minimization routine.

\section{\label{sec5}Conclusion}

In this article, it was proposed to account for two-body interactions in the calculation of partition functions in the canonical ensemble. Based on the work of Faussurier \emph{et al.} relying on Jensen-Feynman (or Gibbs-Bogoliubov) inequality, we showed that, thanks to our recently published recursion relation, it is now possible to use a quadratic reference energy, which takes into account electron-electron interaction inside a given orbital. The required new quantities were presented in a compact form. It was shown, however, using an optimized-exponent reference energy, that a linear reference energy is usually sufficient to achieve a high precision. Finally, we found that the approach of Faussurier \emph{et al.} can be improved by applying Jensen's inequality to the difference between the exponential function and its $(2n-1)^{th}$-order Taylor development.
In this case, the expression of the free energy to be minimized is more complicated, but the quantities (high-order moments) involved can still be obtained from our robust recursion relations.

\section{Appendix A: Averaging process}

When a function $A(\vec p)$ contains only powers of the populations, it is easy to show that its average value $\left\langle A(\vec p)\right\rangle_R$, defined in Eq. (\ref{moyA}), can always be calculated from the knowledge of partition functions, whatever the form of the reference energy. This is due the following relation between binomial coefficients:

\begin{equation}
p\bin{g}{p}=g\bin{g}{p}-g\bin{g-1}{p}.
\end{equation}

This allows one to write, for instance,

\begin{equation}\label{popmoy}
\left\langle p_i\right\rangle_R=g_i\left(1-\frac{U_{Q;N}^R[g^{i}]}{U_{Q;N}^R[g]}\right)
\end{equation}

and

\begin{eqnarray}\label{flucmoy}
&&\left\langle p_i(p_j-\delta_{ij})\right\rangle_R=g_i(g_j-\delta_{ij})\times\nonumber\\
&&\left(1-\frac{U_{Q;N}^R[g^i]}{U_{Q;N}^R[g]}-\frac{U_{Q;N}^R[g^j]}{U_{Q;N}^R[g]}+\frac{U_{Q;N}^R[g^{ij}]}{U_{Q;N}^R[g]}\right),
\end{eqnarray}

where notation $g^{ijk\cdots}$ means that the partition function is evaluated with the degeneracy of orbitals $i$, $j$, $k$, \textit{etc.} reduced by one.

\section{Appendix B: Invariance with respect to a translation of the reference energy}

Let us consider the reference energy

\begin{equation}
E_{R,C}(\vec p)=E_R(\vec p)+C,
\end{equation}

where $C$ is a constant. One has

\begin{eqnarray}
F_{R,C}&=&-\frac{1}{\beta}\ln\left[\sum_{\pinE}G(\vec p)~e^{-\beta E_{R,C}(\vec p)}\right]\nonumber\\
&=&-\frac{1}{\beta}\ln\left[\sum_{\pinE}G(\vec p)~e^{-\beta E_R(\vec p)}\right]+C\nonumber\\
&=&F_R+C.
\end{eqnarray}

Moreover, for any quantity $A$ (see Eq. (\ref{moyA})), one has:

\begin{eqnarray}
\left\langle A\right\rangle_{R,C}&=&\frac{1}{U_{R,C}(\calE)}\sum_{\pinE}A(\vec p)~G(\vec p)~e^{-\beta E_{R,C}(\vec p)}\nonumber\\
&=&\frac{1}{U_R(\calE)}\sum_{\pinE}A(\vec p)~G(\vec p)~e^{-\beta E_R(\vec p)}\nonumber\\
&=&\left\langle A\right\rangle_R.
\end{eqnarray}

Therefore, one obtains

\begin{eqnarray}
\left\langle E-E_{R,C}\right\rangle_{R,C}&=&\left\langle E-E_{R,C}\right\rangle_R\nonumber\\
&=&\left\langle E-E_R\right\rangle_R-C.
\end{eqnarray}

In the same way, one can check easily that the quantities $M_k$ do not depend on $C$. This leads to

\begin{eqnarray}
F^{(2n-1)}&=&F_{R,C}+\left\langle E-E_{R,C}\right\rangle_{R,C}
\nonumber\\
& &-\frac{1}{\beta}\ln\left[1+\sum_{k=2}^{2n-1}(-1)^k\frac{\beta^k}{k!}M_k\right]\nonumber\\
&=&F_R+\left\langle E-E_R\right\rangle_R
\nonumber\\
& &-\frac{1}{\beta}\ln\left[1+\sum_{k=2}^{2n-1}(-1)^k\frac{\beta^k}{k!}M_k\right],
\end{eqnarray}

which means that the quantity $F^{(2n-1)}$ does not depend on $C$, whatever the value of $n$.

\section{Appendix C: Average-atom orbital energies}

There is an error in \cite{fau}, which has not been properly
corrected in the erratum \cite{faue}. In the original paper \cite{fau}, formula
(9) reads:

\begin{equation}
\epsilon_{\alpha}^{ion}=-\epsilon_{\alpha}-\sum_{\gamma}(\left\langle n_{\gamma}\right\rangle_0-
\delta_{\alpha\gamma})\Delta_{\alpha\gamma}.
\end{equation}

In the erratum \cite{faue}, the authors suggest to replace that expression by

\begin{equation}
\epsilon_{\alpha}^{ion}=-\epsilon_{\alpha}-\sum_{\gamma}(\left\langle n_{\gamma}\right\rangle_0-
\delta_{\alpha\gamma}/g_{\alpha})\Delta_{\alpha\gamma},
\end{equation}

which is also wrong. The correct expression is:

\begin{equation}
\epsilon_{\alpha}^{ion}=-\epsilon_{\alpha}-\sum_{\gamma}\left\langle n_{\gamma}\right\rangle_0(1-
\delta_{\alpha\gamma}/g_{\alpha})\Delta_{\alpha\gamma}.
\end{equation}

\clearpage

\begin{table}[t]
\begin{tabular}{|c|c|c|c|}\hline\hline
System & Reference energy $E_R(\vec p)$ & free parameters & degrees of freedom \\ \hline \hline 
A & $\displaystyle\sum_{i=1}^Np_i\epsilon_i+\frac{1}{2}\sum_{i,j=1}^N\left\langle p_i(p_j-\delta_{i,j})\right\rangle V_{ij}$ & none & 0 \\
B & $\displaystyle\sum_{i=1}^Np_i\theta_i$ & $\{\theta_i\}$ & $N$\nonumber\\
C & $\displaystyle\sum_{i=1}^N\left[p_i\theta_i+\frac{1}{2}p_i(p_i-1) \phi_i\right]$ &$\{\theta_i\}$,$\{\phi_i\}$ & $2N$ \\
D & $\displaystyle\sum_{i=1}^N p_i^{\gamma_i} \;\theta_i$ & $\{\theta_i\}$,$\{\gamma_i\}$ & $2N$ \\\hline\hline
\end{tabular}
\caption{Reference energies considered in this paper: A) quadratic part of the energy averaged; B) linear reference energy; C) quadratic reference energy; D) power-law reference energy. The free parameters in cases B, C and D are determined by minimizing the free energy.}\label{tab2bis}
\end{table}

\clearpage

\begin{table}[t]
\begin{tabular}{|c||c|c|}\hline\hline 
& \multicolumn{2}{c|}{$F$} \\
\;\;\;Q\;\;\; & Exact & Exact with $V_{ij}\equiv 0$ \\ \hline\hline
2 & -810.1430 & -829.6907 \\
7 & -1813.4044 & -2187.1193 \\
13 & -1878.6536 & -3115.7014 \\
19 & -914.4643 & -3419.8402 \\
25 & 1377.7764 & -2911.9648 \\\hline\hline
\end{tabular}
\end{table}

\begin{table}[t]
\begin{tabular}{|c||c|c|c|c||c|}\hline\hline 
& \multicolumn{4}{c||}{$F^{(1)}$}&$F^{(3)}$\\
\;\;\;Q\;\;\; & A & B & C & D & B \\ \hline\hline
2 & -809.9468 & -810.1370 & -810.1385 & -810.1386 & -810.1430 \\
7 & -1786.6201 & -1813.1371 & -1813.2003 & -1813.1985 & -1813.4034 \\
13 & -1716.0766 & -1878.0163 & -1878.1632 & -1878.1531 & -1878.6451 \\
19 & -612.4660 &-914.2243 & -914.2825 & -914.2815 & -914.4636 \\
25 & 1474.4451 & 1377.7764 & 1377.7764 & 1377.7764 & 1377.7764 \\\hline\hline
\end{tabular}
\caption{Free energy (eV) for $Q$=2, 7, 13, 19 and 25 electrons in orbitals $(3s_{1/2},\;3p_{1/2},\;3p_{3/2},\;3d_{3/2},\;3d_{5/2},\;4s_{1/2},\;4p_{1/2},\;4p_{3/2})$ for a copper plasma at $T$=100 eV and $\rho$=8.96 g/cm$^3$, calculated using the reference energies A, B, C and D of Table \ref{tab2bis}. The free energy is developed either at the first ($F^{(1)}$) or third ($F^{(3)}$) order of the Jensen-Feynman approach. The second and third columns of the first sub-table above contain respectively the exact values and the values obtained when the interaction energies $V_{ij}$ are artificially cancelled.}\label{tab1}
\end{table}
 
\clearpage

\begin{table}[t]
\begin{tabular}{|c|c|c|c|c|c|c|c|c|}\hline\hline
\;\;\;Q\;\;\; & $3s_{1/2}$ & $3p_{1/2}$ & $3p_{3/2}$ & $3d_{3/2}$ & $3d_{5/2}$ & $4s_{1/2}$ & $4p_{1/2}$ & $4p_{3/2}$ \\ \hline \hline 
2 & 1.001 & 1.001 & 1.001  & 1.002 & 1.002 & 1.023 & 1.026 & 1.028 \\
7 & 1.002 & 1.002 & 1.004 & 1.004 & 1.006 & 1.020 & 1.023 & 1.032 \\
13 & 1.002 & 1.004 & 1.008 & 1.007 & 1.012 & 1.017 & 1.022 & 1.056 \\
19 & 1.001 & 1.001 & 1.006 & 1.003  & 1.012 & 1.026 & 1.032 & 1.133 \\
25 & 1.013 & 1.009 & 1.008 & 1.007 & 1.004 & 1.029 & 1.042 & 1.187 \\\hline\hline
\end{tabular}
\caption{Values of exponents $\gamma_i$ obtained after minimization of the free energy within first-order Jensen-Feynman inequality, using reference energy D: $\sum_{i=1}^Np_i^{\gamma_i}\;\theta_i$.}\label{tab2}
\end{table}

\clearpage

\end{document}